\documentclass[12pt]{article}

\begin{document}

\title{Instanton constraints \\ in supersymmetric gauge theories I.
Supersymmetric QCD}

\author{P. M. Glerfoss, J. Hylsberg, N.K. Nielsen
\footnote{Electronic address: nkn@fysik.sdu.dk}\\
 Physics Department, University of Southern Denmark,\\
Odense, Denmark.}

\date{\today}
\maketitle

\begin{abstract}
A previous analysis of possible constraints of
Yang-Mills instantons in the presence of spontaneous symmetry breaking is
extended to supersymmetric QCD. It is again found that a constraint is
necessary for the gauge field in second and fourth order of
the gauge breaking parameter $v$. While the supersymmetric
zero mode is well behaved to all orders, the lifted
superconformal and quark zero modes show nonpermissible
behaviour, but only at first order in $v$.\\
PACS numbers 11.15.Ðq, 12.60.Jv
\end{abstract}

\section{Introduction}
\label{Int}
Constrained instantons \cite {Affleck} have played an important part in considerations
on nonperturbative effects in quantum theories, notably in supersymmetric 
theories. It
was, however, realized some time ago \cite {NN} that  the class of allowed
constraints is more restricted than originally envisaged since
a scale-fixing constraint is a necessary but not a sufficient
condition for a finite action. This makes it
interesting to examine the original applications of constrained
instantons   in the more restricted class of constraints
allowed according to \cite {NN}. 

In the present paper  instanton
constraints in supersymmetric quantum chromodynamics (SQCD) as originally
considered  in \cite {ADS} are investigated  from this viewpoint. For the
gauge field, where  a brief version of the argument of \cite{NN} is given
for the sake of clarity and completeness, it is again found that a
constraint is necessary in second and fourth order of
the gauge breaking parameter $v$. Our main concern is the
fermionic zero modes, which are determined by iteration in
$v$. The supersymmetric zero mode turns out to be well behaved 
to all orders in $v$, but the superconformal and quark
zero modes, which conspire to develop a nonzero eigenvalue for
$v\neq 0$, show nonpermissible long-distance behaviour
at first order in $v$. The investigation is extended to $N=2$
supersymme\-tric Yang-Mills theory in a separate publication
\cite{NKN}.

Sec.II gives the general setting, with details on the action of
supersymmetric QCD and its extension to Euclidean space. In sec.III the
argument regarding mass corrections of the instanton is recapitulated from
\cite{NN} (with an improved argument for the summation of nextleading
terms at long distances), while sec.IV contains the analysis of the
fermionic zero modes. The results are summarized in the conclusion, and two
appendices contain some technicalities.
 \section{Supersymmetric QCD}

\subsection{The action}
The action of SQCD with gauge group $SU(2)$ is:
\begin{equation}
S_{SQCD}=S_{\rm gauge}+S_{\rm matter}+\tilde{S}_{\rm
matter}.
\end {equation}
Here the gauge field action is
\begin{equation} 
S_{\rm gauge}=\int d^4x(-i(\lambda ^a
_R)^\dagger (\sigma ^\mu D_\mu)^{ab} \lambda
_R^b-\frac{1}{4}F^a_{\mu
\nu }F^{a\mu \nu }
+\frac{1}{2}D^aD^a) \label{LugWZSYM}
\end{equation}
involving the gluon field $A^a_\mu$ with field strength
$F^a_{\mu \nu }$, the gluino field $\lambda _{R}^a$
and the auxiliary field $D^a$, with the covariant derivative $D_\mu=\partial
_\mu-ig[A_\mu,\cdot]$. The matter field action originating from one
chiral superfield 
\begin{eqnarray}
&&S_{\bf matter}
=\int d^4x(-(D^\mu  A)^\dagger D _\mu
A-iq_L^\dagger \bar{\sigma }^{\mu }
D_{\mu }q_L
 +F^\dagger F
\nonumber\\&&
-ig\sqrt{2}A^\dagger \lambda_R^\dagger
 q_L+ig\sqrt{2}q_L^\dagger \lambda _R A-g 
A^\dagger DA)
\label{chiral}
\end{eqnarray}
involves a complex scalar field $A$ and a quark field $q_L$, both in the
fundamental representation of the gauge group, with $D_\mu=\partial
_\mu-igA_\mu $.
$\tilde{S}_{\rm matter}$, originating from the second chiral superfield, has the same
appearance as (\ref{chiral}), but with a new set of "tilded" matter fields 
entering  
according to
\begin{equation}
q_L\rightarrow \tilde{q}_L
\end{equation}
etc. Weyl spinors are used throughout, where with the
conventions of Wess and Bagger \cite{WB} the metric is
$\eta ^{\mu \nu }=(-1,1,1,1)$, while $\sigma
^\mu=(-1,\vec{\sigma}),$ $
\bar{\sigma } ^\mu=(-1,-\vec{\sigma})$, with $\vec{\sigma}$ the
Pauli matrices. Isospin Pauli matrices are denoted $\vec{\tau
}$. In (\ref{LugWZSYM}) and
(\ref{chiral})
\begin{equation}
A _\mu=A _\mu^a\frac{\tau ^a}{2}
\end{equation}
etc.
The action is stationary along the flat
directions where
\begin{equation}
D=0,\hspace{1 mm}A=i\tau ^2\tilde {A}^* \neq 0.
\label{Flashman}
\end{equation}

\subsection {Vainshtein-Zakharov doubling}

In order to continue Majorana spinors to Euclidean space one has
to double the number of components. This is conveniently done by the
method of Vainshtein and Zakharov \cite{Vainshtein}. In the path integral
one defines:
\begin{eqnarray}
&&
\Phi (A^a_\mu, A, \tilde{A})=\int D\lambda ^a_RD(\lambda
^a_R)^\dagger Dq_LDq_L^\dagger D\tilde{q}_LD\tilde{q}_L^\dagger 
\nonumber\\&&
\exp (iS_{{\rm
SQCD, Fermi ,I}})
\label{fermiet}
\end{eqnarray}
where the subscript indicates that only the terms of $S_{{\rm
SQCD}}$ involving fermions are kept.  Here new integration variables
are introduced in accordance with  the Majorana condition 
for the gluino field:
\begin{equation}
\lambda _L^a=- i\sigma ^2(\lambda _R^a) ^*
,\label{Mojorana}
\end{equation}
while for the
the quark fields:
\begin{equation}
q_R=\tau ^2\otimes \sigma ^2 q_L^*, \hspace{1 mm}\tilde {q}_R=\tau
^2\otimes \sigma ^2 (\tilde{q}_L)^*.\label{radaf}
\end{equation}
Expressed in terms of the new integration variables, $\Phi (A^a_\mu, A,
\tilde{A})$ becomes:
\begin{eqnarray}&&
\Phi (A^a_\mu, A, \tilde{A})=\int D\lambda ^a_LD(\lambda
^a_L)^\dagger Dq_RDq_R^\dagger D\tilde{q}_RD\tilde{q}_R^\dagger 
\nonumber\\&&
\exp (iS_{SQCD,
{\rm Fermi ,II}})
\label{fermito}
\end{eqnarray}
with
\begin{eqnarray}
&&S_{SQCD, {\rm Fermi, II}}=\int d^4x(- i(\lambda
^a_L)^\dagger(\bar{\sigma } ^\mu D_\mu )^{ab}\lambda ^b_L 
\nonumber\\&&-
iq_R^\dagger\sigma ^\mu D_\mu q_R
-i\tilde{q}_R^\dagger\sigma ^\mu D_\mu \tilde{q}_R
\nonumber \\&&
-ig\sqrt{2}(\lambda
^a_L)^\dagger A^Ti\tau ^2\frac{\tau ^a}{2}q_R-ig\sqrt{2}
q_R^\dagger\frac{\tau^a}{2}i\tau ^2A^*\lambda ^a_L
\nonumber \\&&
-ig\sqrt{2}(\lambda
^a_L)^\dagger \tilde{A}^Ti\tau ^2\frac{\tau
^a}{2}\tilde{q}_R-ig\sqrt{2}
\tilde{q}_R^\dagger\frac{\tau^a}{2}i\tau ^2\tilde{A}^*\lambda ^a_L).
\end{eqnarray}
From  (\ref{fermiet}) and (\ref{fermito}) is formed
\begin{eqnarray}&&
\Phi ^2(A^a_\mu, A, \tilde{A})=\int D\lambda ^a_LD(\lambda
^a_L)^\dagger D\lambda ^a_RD(\lambda ^a_R)^\dagger Dq_RDq_R^\dagger 
\nonumber\\&&
Dq_LDq_L^\dagger
D\tilde{q}_LD\tilde{q}_L^\dagger D\tilde{q}_RD\tilde{q}_R^\dagger 
\nonumber \\&& 
\exp (iS_{SQCD, {\rm Fermi}}).
\label{fiianden}
\end{eqnarray}
Here all Fermion field integration variables are independent, and
\begin{equation}
S_{SQCD, {\rm Fermi}}=S_{SQCD, {\rm Fermi ,I}}+S_{SQCD,
{\rm Fermi ,II}}.
\end{equation}
This action allows continuation to Euclidean space. This is done by the
following replacements:
\begin{equation}
x^0\rightarrow -ix_4
\end{equation}
and \cite{Espinosa}:
\begin{equation}
\lambda _R\rightarrow \lambda _A,\hspace{1 mm}
\lambda _L\rightarrow\lambda _B,\hspace{1 mm}\lambda
_R^\dagger\rightarrow\lambda _B^\dagger ,\hspace{1 mm}\lambda
_L^\dagger\rightarrow\lambda _A^\dagger
\end{equation}
where the two different Euclidean space Weyl spinors are labeled $A$
and $B$, with the same prescription for the quark fields. The result is:
\begin{eqnarray}&&
\Phi ^2(A^a_\mu, A, \tilde{A})=\int D\lambda ^a_BD(\lambda
^a_A)^\dagger D\lambda ^a_AD(\lambda ^a_B)^\dagger Dq_BDq_A^\dagger 
\nonumber\\&&
Dq_ADq_B^\dagger
D\tilde{q}_BD\tilde{q}_A^\dagger D\tilde{q}_AD\tilde{q}_B^\dagger 
\nonumber \\&& 
\exp (\int d^4xL_{SQCD, {\rm Fermi , Euclid}})
\label{fijianden}
\end{eqnarray}
with
\begin{eqnarray}
&&L_{SQCD, {\rm Fermi,\hspace{1 mm}Euclid}}=
-(\lambda ^a_A)^\dagger(\sigma \cdot
D )^{ab}\lambda ^b_B
%\nonumber\\&&
-(\lambda ^a_B)^\dagger(\bar{\sigma }\cdot
D)^{ab}\lambda ^b_A
\nonumber \\&&
-q_A^\dagger\sigma \cdot D q_B
-q_B^\dagger\bar{\sigma }\cdot D q_A
-\tilde{q}_A^\dagger\sigma \cdot D \tilde{q}_B
-\tilde{q}_B^\dagger\bar{\sigma }\cdot D \tilde{q}_A
\nonumber \\&&
-ig\sqrt{2}A^\dagger(\lambda ^a_B)^\dagger \frac{\tau
^a}{2}q_B+ig\sqrt{2}q^\dagger _A\frac{\tau
^a}{2}\lambda ^a_AA
\nonumber\\&& 
-ig\sqrt{2}\tilde{A}^\dagger(\lambda ^a_B)^\dagger \frac{\tau
^a}{2}\tilde{q}_B
+ig\sqrt{2}\tilde{q}^\dagger _A\frac{\tau
^a}{2}\lambda ^a_A\tilde{A} 
\nonumber \\&&
-ig\sqrt{2}(\lambda
^a_A)^\dagger A^Ti\tau ^2\frac{\tau ^a}{2}q_A-ig\sqrt{2}
q_B^\dagger\frac{\tau^a}{2}i\tau ^2A^*\lambda ^a_B
\nonumber \\&&
-ig\sqrt{2}(\lambda
^a_A)^\dagger \tilde{A}^Ti\tau ^2\frac{\tau
^a}{2}\tilde{q}_A-ig\sqrt{2}
\tilde{q}_B^\dagger\frac{\tau^a}{2}i\tau ^2\tilde{A}^*\lambda ^a_B
\label{lagslut}
\end{eqnarray}
where now $\bar{\sigma }_\mu =(i\vec{\sigma}, 1),\hspace{1 mm}\sigma _\mu
=(-i\vec{\sigma}, 1)$.

\section{Bosonic field equations}

\subsection{General Setup}
\setcounter{equation}{0}

For the moment fermions are ignored. 
The field equations
of the gauge field $A^a_\mu$ and the scalar fields $A$ and
$\tilde{A}$ are from  
(\ref{LugWZSYM}):
\begin{equation}
D _\mu F^a_{\mu \nu } -igA^\dagger \frac{\tau ^a}{2}
\stackrel{\leftrightarrow}{D_\nu}A-ig\tilde{A}^\dagger \frac{\tau ^a}{2}
\stackrel{\leftrightarrow}{D_\nu}\tilde{A}=0
\label{geeq}
\end{equation}
and
\begin{eqnarray}
D^2A=D^2\tilde{A}=0.\label{qeq}
\end{eqnarray}
Equations (\ref{geeq}) and (\ref{qeq}) are simpler than the 
corresponding equations of the Yang-Mills-Higgs system, but a 
similar modification by
means of a constraint is necessary \cite{NN}. It is indicated below  in
some detail how this comes about.

Extremizing the
action with respect to the auxiliary field
$D$ one obtains the solution (\ref{Flashman}).
The following Ansatz is used for the scalar field $A$:
\begin{equation}
A=A(t)u_\tau ;\hspace{1 mm}t=\frac{\rho ^2}{x^2} 
\label{Ansatz}
\end{equation}
with  $\rho $ the scale of
the instanton and $u_\tau $ a constant unit isospinor.
The gauge field is in the singular gauge supposed to have the form:
\begin{equation}
A^a_{\mu }
=-\frac{1}{g}\bar{\eta }^a_{\mu \nu }\partial _{\nu }\log
\alpha (t),
\label{ansatz}
\end{equation}
with $\bar{\eta}^a_{\mu \nu }$ the standard 't Hooft symbol.
Then (\ref{geeq}) and (\ref{qeq}) are, expressed in terms of the functions
$A(t)$ and $\alpha (t)$:
\begin{eqnarray}
  \frac{d}{dt}\left(\alpha^{-3}t^3\frac{d^2\alpha}{dt^2}\right)
  =\frac{\rho^2g^2A^2}{4}\alpha^{-3}\frac{d\alpha}{dt}
\label{ali1}
\end{eqnarray}
and
\begin{eqnarray}
  \alpha^2\frac{d^2A}{dt^2}-\frac{3}{4}\left(\frac{d\alpha}{dt}\right)^2A
  =0.  \label{ali2}
\end{eqnarray}

\subsection{Iteration}

The equations (\ref{ali1}) and (\ref{ali2}) are solved iteratively.
In the two lowest  orders  
\begin{equation}
\alpha _0=1+t;\hspace{1 mm}A_1=\frac{v}{\sqrt{1+t}}
\label{Aen}
\end{equation}
where the constant $v$, in terms of which the expansion is carried out,  is undetermined,
and the subscript here and henceforth denotes the order of $v$.

Finiteness of the action leads according to the
analysis of \cite{NN} to restrictions of the form of the prepotential
$\alpha $ both at  small and large values of $t$. At small values of $t$
the leading terms of $\alpha $ should  conspire to the modified
Bessel funtion $K_1$:
\begin{equation}
\alpha=\alpha_0+\alpha_2+\alpha_4+\cdots\simeq 1+\rho gv\sqrt t
K_1(\frac{\rho gv}{\sqrt t}), \label{an1}
\end{equation}
thus ensuring exponential falloff of the
gauge field. For large values of $t$ finiteness of the action
requires $\alpha $ to grow at most like $\log t$ to all orders
in $v$.
The properties of modified Bessel functions relevant for the present
investigation are listed in Appendix B and from (\ref{watson}) follows
\begin{equation}
\alpha _2\simeq \frac{\rho ^2g^2v^2}{4}
(\log \frac{\rho ^2g^2v^2}{4t}+2\gamma -1)
\label{Besselto}
\end{equation}
valid near $t=0$.

The equation determining $\alpha _2$ is according to
(\ref{ali1}):
\begin{eqnarray}
\frac{d}{dt}\left(\frac{t^3}{(1+t)^3}\frac{d^2\alpha_2}{dt^2}\right)
=\frac{\rho^2g^2v^2}{4}\frac{1}{(1+t )^4}.
\label{secondorder}
\end{eqnarray}
This equation has at $t\rightarrow 0$ ($x\rightarrow \infty $) the
form
\begin{equation}
\frac{d}{dt}t^3\frac{d^2\alpha_2}{dt^2}
\simeq\frac{\rho^2g^2v^2}{4}
\end{equation}
of which  (\ref{Besselto}) is a solution, and for $t\rightarrow
\infty$ ($x\rightarrow 0$) the form of (\ref{secondorder}) is:
\begin{equation}
\frac{d^3\alpha _2}{dt^3}\simeq O(t^{-4})
\label{urukhai}
\end{equation}
that is consistent with $\alpha _2$ being bounded in this limit. Thus on
this level of analysis a constraint is not required. 

Solving (\ref{secondorder}) one gets:
 \begin{equation}
\frac{d^2\alpha_2}{dt^2}=-
\frac{\rho^2g^2v^2}{12t^3}+c_{2;1}(\frac{1+t}{t})^3
\label{aral}
\end{equation}
with $c_{2;1}$ a constant of integration. The most singular part of $\alpha_2$ for
$t\rightarrow 0$ is then
\begin{equation}
\alpha _2\simeq -\frac{1}{2t}(\frac{\rho^2g^2v^2}{12}-c_{2;1}).
\end{equation}
Such a term is not permitted, so the value of  $c_{2;1}$ is
fixed: 
\begin{equation}
c_{2;1}=\frac{\rho^2g^2v^2}{12}.
\end{equation}
This, however, upsets the asymptotic estimate (\ref{urukhai}); it is
replaced by:
\begin{equation}
\frac{d^2\alpha_2}{dt^2}\simeq c_{2;1}(\frac 3t+1)
\label{juanjuan}
\end{equation}
leading to the solution 
\begin{equation}
\alpha_2=\frac{\rho^2g^2v^2}{4}(
  -\log t+t\log t-t+\frac{1}{6}t^2)+
c_{2;2}t+c_{2;3}
\label{onoghur}
\end{equation}
where $c_{2;2}$ and $c_{2;3}$ are new integration constants, with
$c_{2;2}=\frac{\rho^2g^2v^2}{4}$ while
$c_{2;3}$ is fixed by comparison with (\ref{Besselto}).

The terms 
$$
\frac{\rho^2g^2v^2}{4}(t\log t+\frac{1}{6}t^2)
$$ 
of (\ref{onoghur}) have to be eliminated
for a finite-action solution.
This can be accomplished by modifying (\ref{secondorder}) to:
\begin{equation}
\frac{d}{dt}\left(\frac{t^3}{(1+t)^3}\frac{d^2\alpha_2}{dt^2}\right)
  =\frac{\rho^2g^2v^2}{4}\frac{1}{(1+t)^4}-\frac{\rho^2g^2v^2}{2}\frac{t}{(1+t)^4}
\label{alfa2modmod}
\end{equation}
where the extra term is introduced by a constraint. 
The solution of (\ref{alfa2modmod}) is  (\ref{Besselto}), now with
equality sign.

At third order $A_3$ is according to
(\ref{ali2}) determined by:
\begin{eqnarray}
&&(1+t)^2\frac{d^2A_3}{dt^2}-\frac{3}{4}A_3
%\nonumber\\&&
  =\frac{3}{2}v\sqrt{1+t}
\frac{d}{dt}\frac{\alpha _2}{1+t}
\end{eqnarray}
 with the following solution, with
appropriate choices of the two integration constants:
\begin{eqnarray}&&
A_3
 =-\frac 12\frac{v}{\sqrt{1+t}}\frac{\alpha _2}{1+t}
\nonumber\\&&-\frac{\rho ^2g^2 v
^2}{8}\frac{v}{\sqrt{1+t}}
  \left (t-(1+t)^2\log\frac{1+t}{t}
  \right ).
\label{tildeftre}
\end{eqnarray}

At fourth order the equation determining $\alpha _4$ must also be modified
by a constraint. It is according to
(\ref{ali1}):
\begin{eqnarray}&&
\frac{d}{dt}\left(\frac{t^3}{(1+t)^3}\frac{d^2\alpha_4}{dt^2}
  -3\alpha_2\frac{t^3}{(1+t)^4}\frac{d^2\alpha_2}{dt^2}\right)
\nonumber \\
  &&=\frac{\rho ^2g ^2v^2}{4}\left(\frac{d}{dt}
    \left(\frac{\alpha_2}{(1+t)^4}\right)+\frac{\frac{2A_3\sqrt{1+t}}{v}+\frac{\alpha
_2}{1+t}}{(1+t)^4}\right).
\label{alpha4}
\end{eqnarray}
The first integration yields by (\ref{tildeftre}) 
the result:
\begin{eqnarray}&&
\frac{t^3}{(1+t)^3}\frac{d^2\alpha_4}{dt^2}=3\alpha_2\frac{t^3}{(1+t)^4}
\frac{d^2\alpha_2}{dt^2}
+\frac{\rho ^2g^2v^2}{4}\frac{\alpha_2}{(1+t)^4}
\nonumber \\
&&-(\frac{\rho ^2g^2v ^2}{4})^2
\left(-\frac{t}{1+t}\log (1+\frac{1}{t})+\frac{1}{1+t}
-\frac{1}{2}\frac{1}{(1+t)^2}\right .
\nonumber\\&&
\left .+\frac{1}{3}\frac{1}{(1+t)^3}\right)
+c_{4;1}
\label{susquehannah}
\end{eqnarray}
with $c_{4;1}$ a new integration constant. From (\ref{susquehannah}) one
finds the most singular terms of $\alpha _4$ for
$t\rightarrow 0 $:
\begin{eqnarray}&&
\alpha_4\simeq (\frac{\rho^2 g^2v^2}{4})^2
  (\log\frac{\rho ^2g^2v ^2}{4t}+2\gamma-\frac 52)
  \frac{1}{2t}
\nonumber\\&&+(c_{4;1}-\frac
56(\frac{\rho^2g^2v^2}{4})^2)\frac{1}{2t}.
\end{eqnarray}
This should agree with  terms of order
$v^4$ in $\rho gv\sqrt t
K_1(\frac{\rho gv}{\sqrt t})$  found from (\ref{watson}), and hence
$c_{4;1}$ is fixed:
\begin{equation}
c_{4;1}=\frac
56(\frac{\rho^2g^2v^2}{4})^2\neq 0.
\end{equation}

Letting next $t\rightarrow\infty$ in (\ref{susquehannah}) and
disregarding all terms on the right hand side which vanish in
this limit one gets
\begin{equation}
\frac{d^2\alpha_4}{dt^2}\simeq c_{4;1}(\frac{3}{t}+1)
\label{c41}
\end{equation}
that is similar to (\ref{juanjuan}) and gives rise to the same problem. Thus (\ref{alpha4})
should have an additional term on the right hand side
\begin{equation}
-c_{4;1}\frac{6t}{(1+t)^4}
\end{equation}
(cf. (\ref{secondorder}) and (\ref{alfa2modmod})). 

A complete determination of $\alpha _4$ is now possible, and
the result is similar to the corresponding result for the
Yang-Mills-Higgs system reported in \cite{NN}.

\subsection{Constraint term in the gauge field equation} 
The required modifications of (\ref{secondorder}) and (\ref{alpha4}) are
achieved by replacing  (\ref{ali1})  with:
\begin{equation}
\alpha ^2\frac{d}{dt}\left(\alpha
^{-3}t^3\frac{d^2\alpha}{dt^2}\right)+c\frac{6t}{(1+t)^2}
=\frac{\rho^2g^2A^2}{4 }\alpha^{-1}\frac{d\alpha}{dt}
\label{alixx1}
\end{equation}
where
\begin{equation}
c=c_{2;1}+c_{4;1}.
\end{equation}
 Thus the gauge field equation (\ref{geeq}) is modified to:
\begin{eqnarray}&&
D _\mu F^a_{\mu \nu }+\frac{c}{g}
 \bar{\eta}^a_{\nu
\lambda }x_\lambda\frac{48   \rho ^2}{x^2(\rho ^2+x^2)^2}
\nonumber\\&& -igA^\dagger \frac{\tau ^a}{2}
\stackrel{\leftrightarrow}{D_\mu}A-ig\tilde{A}^\dagger \frac{\tau ^a}{2}
\stackrel{\leftrightarrow}{D_\mu}\tilde{A}=0.
\label{qeeqo}
\end{eqnarray}
The extra term in the field equation is provided by a source term
in the action that in its turn is obtained from a constraint.
The scalar field equation (\ref{qeq}) requires no modification. 

\subsection{Asymptotic Estimates}

The functions $\alpha _n,\hspace{1 mm}n\neq 0,$ and $\sqrt t A_n$
will for 
$t\rightarrow \infty $ $ (x\rightarrow 0)$ in each order of
$v$ diverge at most logarithmically.  This follows by induction 
from  (\ref{ali1}) (supplemented with the constraint term)
and (\ref{ali2}) by the proof outlined in
\cite{NN}.

In the opposite limit, $t\rightarrow 0$, the low order calculations
lead to the following estimate:
\begin{equation}
\alpha _n\propto t^{1-\frac{n}{2}}
\end{equation}
which is seen from (\ref{ali1}) to be consistent to all orders. The leading term is thus for
$n>4$ more singular than $\frac 1t$, and will therefore not mix with the terms
arising from the integration constants, which here can be taken equal to zero.  No
new constraint terms are therefore required.

The leading terms of $\alpha $ and $A$ in this limit, obtained by summing leading terms of
$\alpha $ and $A$ to all orders in $v$, are obtained by a series expansion in $\rho $ with
$x$ kept fixed, or alternatively by a double series expansion in $\rho $ and
$\sqrt t$. The order in this new expansion is indicated by a
superscript.
Leading terms  are
\begin{equation}
A^{(0)}=v, \hspace{1 mm}\alpha ^{(0)}=1
\label{gigantneo}
\end{equation}
and
\begin{equation}
A^{(2)}(t)=-\frac{vt}{2}, 
\alpha ^{(2)}(t)= \rho gv\sqrt tK_1(\frac{\rho gv}{\sqrt
t})
\label{superarnie}
\end{equation}
(cf. (\ref{an1})).

The nextleading terms can as shown in \cite{NN} be summed by
Green's function techniques. We give below an
improved version of this argument.

 (\ref{ali1}) and (\ref{ali2}) are in fourth
order of the expansion in $\rho $ and $\sqrt t$:
\begin{eqnarray}&&
\frac{d^2}{dt^2}\frac{d\alpha
^{(4)}}{dt}+\frac{3}{t}\frac{d}{dt}\frac{d\alpha
^{(4)}}{dt}-\frac{\rho ^2g^2v^2}{4t^3}\frac{d\alpha
^{(4)}}{dt}
\nonumber\\&&=\frac{\rho ^2g^2v}{4t^3}(3v\alpha
^{(2)}+2A^{(2)})\frac{d\alpha ^{(2)}}{dt}
\label{alwyn}
\end{eqnarray}
and
\begin{eqnarray}
  \frac{d^2A^{(4)}}{dt^2}=\frac{3}{4}
\left(\frac{d\alpha^{(2)}}{dt}\right)^2v.
    \label{alsil2}
\end{eqnarray}
The constraint gives according to
(\ref{alfa2modmod}) an extra term on the right hand
side of (\ref{alwyn}):
\begin{equation}
-\frac{\rho ^2g^2v^2}{2}\frac {1}{t^2}.
\label{alyozha}
\end{equation}
This means that on the
right hand side of (\ref{alwyn}), when expanded in powers of $v$, all $O(t^{-2})$ terms
cancel out, such that the lowest order term is $O(t^{-3})$.

The solution of (\ref{alsil2}) is found by quadrature, where
from (\ref{Aen}):
\begin{equation}
A^{(4)}(t)=\frac 38vt^2+\cdots.
\end{equation}
The solution is
\begin{equation}
A^{(4)}(t)=\frac 34v\int
_0^tdt'(t-t')\left(\frac{d\alpha^{(2)}(t')}{dt'}\right)^2
\end{equation} 
showing exponential falloff for $t\rightarrow 0$, in contrast to
$A^{(2)}(t)$.

(\ref{alwyn}) is a special case of the following equation dealt with in App.B:
\begin{equation}
(\frac{d^2}{dt^2}+\frac{1-n}{t}\frac{d}{dt}-\frac{\rho
^2m^2}{4t^3})f(t)=J(t)
\label{crocodyl}
\end{equation}
where $n$ is an integer, with the general solution
\begin{eqnarray}&&
f(t)=- \int
_{t_0}^tdt'(f_{1}(t)f_{2}(t')-f_{1}(t')f_{2}(t))W^{-1}(t')J(t')
\nonumber\\&&
+C_1f_1(t)+C_2f_2(t)
\label{minotaur}
\end{eqnarray}
with $C_1$ and $C_2$ integration constants. Here $f_1(t)$ and $f_2(t)$, given in
(\ref{Bessel}), are indepent solutions of the corresponding homogeneous
equation with the Wronskian 
\begin{equation}
W(t)=f_{1}(t)\frac{df_{2}(t)}{dt}-f_{2}(t)\frac{df_{1}(t)}{dt}\propto
t^{n-1}.
\label{woeronski}
\end{equation}

The asymptotic form of (\ref{minotaur}) after
summation over all orders of $v$ can be found from  (\ref{ask}), and it turns out that the term of $\alpha
^{(4)}$ of lowest order in $v$ restricts the solution
in such a way that  exponential increase of $\alpha ^{(4)}$ for
$t\rightarrow 0$ is ruled out.

Inserting  the lowest powers from (\ref{whitaker}) and (\ref{watson}) into
(\ref{Bessel}) (with $n=-2$)
one obtains the following lowest order terms in the power series
expansions of the two independent solutions
$f_1(t)$ and $f_2(t)$  of the homogeneous
version of (\ref{alwyn}):
\begin{equation}
f_1(t)\simeq \frac {\rho ^4}{8t^2};\hspace{1 mm}f_2(t)\simeq
\frac{2}{g^4v^4}.
\label{estmate}
\end{equation}
Estimating the two integral terms of (\ref{minotaur}) for the
present case near
$t\simeq 0$ one finds
$W^{-1}(t)J(t)=O(t^0)$, where $J(t)$ now denotes the right hand side of
(\ref{alwyn}). Thus
\begin{equation}
\int
_{t_0}^tdt'f_{1}(t')W^{-1}(t')J(t')=O(t^{-1})
\end{equation}
and
\begin{equation}\int
_{0}^tdt'f_{2}(t')W^{-1}(t')J(t')=O(t)
\end{equation}
and the two integral terms of (\ref{minotaur}) are both
$O(t^{-1})$ in this case.

The solution $\frac{d\alpha
^{(4)}}{dt}$ of (\ref{alwyn}) is according to the general 
estimate
$\alpha _n=O(t^{2-\frac n2})$ for nextleading terms 
of order $t^{-1}$ at lowest order in $v$; this is also a consequence of
(\ref{susquehannah}). The only part of
(\ref{minotaur}) that is more singular is according to the result of the
last paragraph the
term $C_1f_1(t)$.  Thus 
$C_1=0$ and (\ref{minotaur}) becomes 
\begin{eqnarray}&&
f(t)=- f_{1}(t)\int
_{0}^tdt'(t)f_{2}(t')W^{-1}(t')J(t')
\nonumber\\&&
+f_{2}(t)\int
_{t_0}^tdt'f_{1}(t')W^{-1}(t')J(t')
+C_2f_2(t).
\label{mixotaur}
\end{eqnarray}
In (\ref{mixotaur})
all terms have exponential falloff for $t\rightarrow 0$ by
(\ref{ask}) and (\ref{maxotaur}), provided the source function
$J(t)$ has this property. This is not the case for the term
(\ref{alyozha}) arising from the constraint; however, as shown
in \cite{NN} an exponential factor may be included in the
constraint to ensure exponential falloff of the nextleading
terms arising from the constraint.

The iteration procedure outlined here can obviously be continued to
higher orders.

\section{The supersymmetric zero mode}
\subsection{General setup}
\setcounter{equation}{0}

From (\ref{lagslut}) one obtains the following equations for the gluino
zero modes:
\begin{equation}
(\sigma \cdot
D )^{ab}\lambda ^b_B+ig\sqrt{2}A^Ti\tau ^2\frac{\tau
^a}{2}q_A
+ig\sqrt{2}\tilde{A}^T\frac{\tau
^a}{2}\tilde{q}_A=0,  \label{lambada}
\end{equation}
\begin{equation} 
\bar{\sigma }\cdot D q_A+ig\sqrt{2}\lambda ^a_B\otimes \frac{\tau
^a}{2}i\tau ^2A^*=0,  \label{qu}
\end{equation}
and
\begin{equation}
\bar{\sigma }\cdot D \tilde{q}_A+ig\sqrt{2}\lambda ^a_B\otimes \frac{\tau
^a}{2}i\tau ^2\tilde{A}^*=0. \label{tildequ}
\end{equation}

In the following  boson background
fields are those calculated in the previous subsection. $u_\sigma$ is
a constant unit two-spinor, and $u_\tau $ the corresponding isospinor
introduced in (\ref{Ansatz}), and the quark components of the
zero modes are written as direct products of isospinors and
spinors. 

With the Ans\"{a}tze, relevant for the supersymmetric zero modes in the
singular gauge of the instanton field:
\begin{equation}
\lambda ^a_B(x)=f(t)\bar{\sigma }\cdot x\sigma ^a\sigma \cdot
xu_\sigma ,
\end{equation}
\begin{equation}
q_A(x)=i\phi (t)\tau \cdot x\bar{\tau}_\mu i\tau ^2u_\tau \otimes\sigma
_\mu u_\sigma+i\psi (t)i\tau ^2u_\tau\otimes \sigma \cdot xu_\sigma
\end{equation}
and
\begin{equation}
\tilde{q}_A(x)=i\phi (t)\tau \cdot x\bar{\tau}_\mu u_\tau
\otimes\sigma _\mu u_\sigma
+i\psi (t)u_\tau\otimes \sigma
\cdot xu_\sigma
\end{equation}
the equations (\ref{lambada}), (\ref{qu}) and
(\ref{tildequ}) are reformulated by means of the identities listed in
App.\ref{Ydyntytyt} into a set of coupled first order differential
equations:
\begin{equation}
6f(t)-2t\frac{df(t)}{dt}-4t\alpha
(t)^{-1}\frac{d\alpha (t)}{dt} f(t)=-g\sqrt 2A(t)\phi (t),
\label{harlequin}
\end{equation}
\begin{eqnarray}&&
4(\phi (t)+\psi
(t))-2t\frac{d\psi (t)}{dt}
%\nonumber\\&&
+t\alpha ^{-1}(t)\frac{d\alpha
(t)}{dt}(-4\phi (t)+\psi (t))
\nonumber\\&&=\frac{g}{\sqrt 2}\frac{\rho ^2}{t}A(t)f(t)
\label{columbine}
\end{eqnarray}
and
\begin{equation}
%&&
 2\frac{d\phi (t)}{dt}+\alpha
^{-1}(t)\frac{d\alpha (t)}{dt}(-\phi (t)+\psi (t))
=\frac{g}{\sqrt 2}\frac{\rho ^2}{t^2} A(t)f(t).
\label{Pierrot}
\end{equation}

\subsection{Iteration: first four orders}

The set of coupled  equations (\ref{harlequin}), (\ref{columbine})
and (\ref{Pierrot}) is solved by iteration in the parameter $v$, with even
orders for the function $f$ and odd orders for the functions $\phi$ and
$\psi $. The order of $v$ is as in the previous section indicated by a
subscript.

To zeroth order (\ref{harlequin}) implies:
\begin{equation}
6f_0(t)-2t\frac{df_0(t)}{dt}-\frac{4t}{1+t}f_0(t)=0
\end{equation}
i.e.
\begin{equation}
f_0(t)=\frac{t^3}{(1+t)^2}\label{ursus}
\end{equation}
where a proportionality factor was set equal to unity.

At first order  (\ref{columbine}) and (\ref{Pierrot}) reduce to:
\begin{eqnarray}&&
4(\phi _1(t)+\psi _1
(t))-2t\frac{d\psi _1(t)}{dt}+\frac{t}{1+t}(-4\phi _1(t)+\psi _1(t))
\nonumber\\&&=\frac{\rho ^2gv}{\sqrt 2}\frac{t^2}{(1+t)^{\frac 52}}
\label{columbo}
\end{eqnarray}
and
\begin{equation}
2\frac{d\phi _1(t)}{dt}+\frac{1}{1+t}(-\phi _1(t)+\psi _1(t))=
\frac{\rho ^2gv}{\sqrt 2}\frac{t}{(1+t)^{\frac 52}}
\label{Pierrotlalune}
\end{equation}
with the solution
\begin{equation}
\phi _1(t)=\frac{\rho^2gv\sqrt{2}}{8}
\frac{t^2}{(1+t)^{\frac{3}{2}}}, \psi _1(t)=0.
\label{ursula}
\end{equation}

At second order the following equation arises from
(\ref{harlequin}):
\begin{eqnarray}
&&6f_2(t)-2t\frac{df_2(t)}{dt}-\frac{4t}{1+t}f_2(t)
=-\frac{\rho ^2g^2v^2}{4}
\frac{t^2}{(1+t)^{2}}
\nonumber \\&& 
+\frac{4t^4}{(1+t)^2}(\frac{d}{dt}\frac{\alpha
_2(t)}{1+t})
\end{eqnarray}
with the solution
\begin{equation}
f_2(t)=-\frac{\rho
^2g^2v^2}{8}\frac{t^2}{(1+t)^2}-\frac{2t^3\alpha
_2(t)}{(1+t)^3}.
\label{belsazar}
\end{equation}

At third order the functions $\phi _3(t)$ and $\psi _3(t)$ are 
determined through the coupled equations:
\begin{eqnarray}
&&\frac{4\phi _3(t)}{(1+t)^\frac12}-2t^3(1+t)\frac{d}{dt}\frac{\psi 
_3(t)}{t^2(1+t)^\frac12}
\nonumber\\&&
=gv\sqrt{2}\frac{\rho ^4g^2 v
^2}{16}(t^2\log\frac{1+t}{t}-t)
\nonumber \\&&
-\frac{\rho^2gv\sqrt{2}}{4}\left(\frac{5t^2}{(1+t)^{3}}+\frac{2t^3}{(1+t)^{3}}\right)
\alpha _2(t)\label{fyrstecoupled}
\end{eqnarray}
and
\begin{eqnarray}
&&
2(1+t)\frac{d}{dt}\frac{\phi
_3(t)}{(1+t)^{\frac12}}+\frac{\psi _3(t)}{(1+t)^\frac12}
\nonumber\\&&
=gv\sqrt{2}\frac{\rho ^4g^2 v
^2}{16}(t\log\frac{1+t}{t}-1)
\nonumber \\&&
-\frac{\rho^2 gv\sqrt{2}}{4}
\left(\frac{5t}{(1+t)^{3}}+\frac{1}{2}\frac{t^2}{(1+t)^{3}}\right)\alpha
_2(t)
\nonumber \\&&
+3gv\sqrt{2}\frac{\rho
^4g^2v^2}{32}\frac{t}{(1+t)^{2}}.\label{annencoupled}
\end{eqnarray}

In contrast to first order, both functions $\phi _3$ and
$\psi _3$ are nonvanishing; after some computation 
the following expressions are found: 
\begin{eqnarray}
&&\frac{\phi_3(t)}{(1+t)^\frac12}=-\frac{\rho ^2gv\sqrt{2}}{16}
\left(\frac{2t^3}{(1+t)^3}+\frac{5t^2}{(1+t)^3}\right)\alpha _2(t)
\nonumber \\&&
-gv\sqrt{2}\frac{\rho
^4g^2v^2}{48}((1+\frac{t}{2(1+t)})\log
(1+t)-\frac 52t)
\label{alcazar}
\end{eqnarray}
and
\begin{eqnarray}
&&\frac{\psi
_3(t)}{(1+t)^\frac12}=-gv\sqrt{2}\frac{\rho^4g^2v^2}{16}(t\log
\frac{1+t}{t}-1)
\nonumber\\&&
\left (\frac{1}{2}t+\frac{1}{6}\frac{t^2}{1+t}\right)
+\frac{gv\sqrt{2}}{1+t}\frac{\rho^4g^2v^2}{48}
(\log (1+t)-t).
\label{alcatraz}
\end{eqnarray}

For $t\rightarrow \infty$ the function
$\frac{\phi_{3}(t)}{(1+t)^\frac12}$ and 
$\psi_{3}(t)(1+t)^\frac12$
 only have  logarithmic growth.
Thus the zero mode remains square integrable in this order, and no constraint
is required to ensure acceptable asymptotic behaviour.

\subsection {Ultraviolet asymptotic behaviour of the
supersymmetric zero mode}

In a similar way as in the bosonic case, it is proved by induction on
the basis of (\ref{harlequin}), (\ref{columbine}) and
(\ref{Pierrot}) that 
$\frac{\phi (t)}{\sqrt t}$, $\sqrt t\psi (t)$ and $f(t)$  at most have
logarithmic growth for $t\rightarrow \infty $ at each order in $v$, except
$f_0$ which grows linearly. 
Assuming that the above-mentioned estimates hold to orders less than $n$,
one sees immediately from (\ref{harlequin}) that $f _n$ is $O(t^0)$ for
$t\rightarrow \infty$. From  (\ref{columbine}) one gets:
\begin{equation}
\frac{4\phi _n(t)}{1+t}+4\psi _n(t)-2t\frac{d\psi _n(t)}{dt}+\frac{t\psi _n(t)}{1+t}
\propto t^{-\frac 12}
\end{equation}
and from (\ref{Pierrot}):
\begin{equation}
2\frac{d\phi_n(t)}{dt}-\frac{\phi _n(t)}{1+t}+\frac{\psi _n(t)}{1+t}\propto t^{-\frac
32}
\end{equation} 
establishing the estimates to order $n$ also for the functions $\phi $
and $\psi $. Thus the estimate holds to all orders.

\subsection{Infrared asymptotic behaviour of the
supersymmetric zero mode}

A power counting argument is carried out to determine the leading terms
for $t\rightarrow 0$ of $f$, $\phi$ and $\psi$ in each order of
$v$. From the leading order results, extracted from 
(\ref{ursus}), (\ref{ursula}), (\ref{belsazar}), (\ref{alcazar}) and
(\ref{alcatraz}),
one is lead to the hypothesis
\begin{equation}
f_{ n}\propto t^{3-\frac{n}{2}}, \hspace{1 mm}
  \phi _{ n}\propto t^{\frac{5}{2}-\frac{n}{2}}
\label{duncan}
\end{equation}
while $\phi _n+\psi _n$ is subleading, except at $n=1$.

The leading terms summed to all orders in $v$ are again  obtained by a
systematic double power expansion in $\rho $ and $\sqrt t$, where the
leading order estimates of the bosonic functions listed in
(\ref{gigantneo}) and (\ref{superarnie}) are combined with leading terms
of $f$ and $\phi$, which are of sixth order in the double expansion of
$\rho$ and $\sqrt t$ and according to (\ref{ursus}) and (\ref{ursula})
have the following lowest order terms in the expansion of
$v$:
\begin{equation}
f^{(6)}(t)=t^3+\cdots,\hspace{1 mm}
\phi ^{(6)}(t)=\frac{g\sqrt 2v\rho ^2}{8}t^2+\cdots .
\label{ichtysaur}
\end{equation}
They are solutions of\begin{equation}
6f^{(6)}(t)-2t\frac{df^{(6)}}{dt}=-g\sqrt 2v\phi ^{(6)}(t)
\label{macbeth}
\end{equation}
and
\begin{equation}
\frac{2t^2}{\rho ^2}\frac{d\phi ^{(6)}}{dt}=\frac{gv}{\sqrt 2}f^{(6)}(t)
\label{alyceford}
\end{equation}
following from (\ref{harlequin}), (\ref{columbine}) and
(\ref{Pierrot}), where it also is used that $\phi +\psi $ only has an
eighth order term.
Combining (\ref{macbeth}) and (\ref{alyceford}) one gets:
\begin{equation}
\frac{d^2f^{(6)}(t)}{dt^2}-\frac{2}{t}\frac{df^{(6)}(t)}{dt}
-\frac{g^2v^2\rho^2}{4t^3}f^{(6)}(t)=0
\label{LLL}
\end{equation}
and
\begin{equation}
\frac{d^2\phi ^{(6)}(t)}{dt^2}
-\frac{1}{t}\frac{d\phi ^{(6)}(t)}{dt}
-\frac{g^2v^2\rho^2}{4t^3}\phi ^{(6)}(t)=0
\label{QQQ}
\end{equation}
which are special cases of (\ref{minotaur}) and have the following solutions, obtained by
combination of (\ref{Bessel}), the lowest order terms of the modified Bessel functions
(\ref{whitaker}) and (\ref{watson}), and finally
(\ref{ichtysaur}):
\begin{eqnarray}&&
f^{(6)}(t)=\frac{\rho ^6}{8}(\frac{gv\sqrt t}{\rho})^3K_3(\frac{\rho gv}{\sqrt
t})),\\&&\phi ^{(6)}(t)=\frac{g\sqrt 2v\rho ^6}{16}(\frac{gv\sqrt
t}{\rho})^2K_2(\frac{\rho gv}{\sqrt t}).
\label{cavarichmond}
\end{eqnarray}
It is an important point that the modified Bessel functions $I_2$ and $I_3$ do not occur
here; as a consequence the two solutions show exponential decrease for
$t\rightarrow 0$ ($t\rightarrow \infty$). 

We show explicitly how  the term added to the first logarithmic part
of $f$ and $\phi$ can be chosen
freely as integration constants, thus excluding the presence  $I_2$ and
$I_3$ in the solutions.  We thus return to the power series expansion in $v$, where
logarithmic terms occur in $\phi _5$ and $f_{ 6}$. Hence we are led to consider separately: 
\begin{equation}
-2t^4\frac{d}{dt}\frac{f_4(t)}{t^3}\simeq -g\sqrt{2}v\phi_{3}(t),
\end{equation}
\begin{equation}
\frac{d\phi_5
(t)}{dt}\simeq \frac{g\sqrt{2}v\rho^2}{4t^2}f_{4}(t)
\end{equation}
and
\begin{equation}
-2t^4\frac{d}{dt}\frac{f_6(t)}{t^3}\simeq -g\sqrt{2}v\phi _{5}(t)
\end{equation}
with, according to (\ref{alcazar}):
\begin{equation}
\phi_3(t)\simeq -g\sqrt{2}v\frac{\rho ^4g^2v^2}{32}t.
\end{equation}
The solutions of these equations are first
\begin{equation}
f_4(t)\simeq \frac{\rho ^4g^4v^4}{64}t
\end{equation}
and
\begin{equation}
\phi_5(t)\simeq -g\sqrt{2} v \frac{\rho ^4g^4v^4}{256}(\log
\frac{\rho ^2m^2}{4t}+2\gamma -\frac 32)
\end{equation}
where an integration constant is chosen in accordance with
(\ref{watson}).
Finally $f_6(t)$ is determined by:
\begin{equation}
\frac{d}{dt}\frac{f_6(t)}{t^3}\simeq -\frac{\rho ^4g^6v^6}{256}
(\log
\frac{\rho ^2m^2}{4t}+2\gamma -\frac 32)\frac{1}{t^4}
\end{equation}
whence
\begin{equation}
f_6(t)\simeq \frac{\rho ^4g^6v^6}{768}
(\log
\frac{\rho ^2m^2}{4t}+2\gamma -\frac 32-\frac 13)
\end{equation}
in exact ageement with (\ref{watson}). Thus correct Bessel function
behaviour of both the quark and gluino functions is ensured once
the integration constant in $\phi_5$ has been chosen properly.

Nextleading terms are of eighth order in $\rho $ and $\sqrt t$ and are
according to (\ref{harlequin}), (\ref{columbine}) and (\ref{Pierrot})
determined by:
\begin{eqnarray}&&
\frac{d^2f^{(8)}(t)}{dt^2}-\frac 2t\frac{df^{(8)}(t)}{dt}-\frac{g^2v^2\rho^2}{4t^3
}f^{(8)}(t)
\nonumber\\&& 
=\frac{g\sqrt 2v\rho^2}{4t^3}(\frac{g}{\sqrt
2}A^{(2)}f^{(6)}(t)+\frac{3t^2}{\rho ^2}\frac{d\alpha ^{(2)}(t)}{dt}\phi
^{(6)}(t))
\nonumber\\&&
-\frac {1}{2t}\frac
{d}{dt}(-g\sqrt 2vA^{(2)}\phi ^{(6)}+4t\frac{d\alpha ^{(2)}}{dt}f^{(6)}(t)),
\label{louellen}
\end{eqnarray}
\begin{eqnarray}&&
\frac{d^2\phi ^{(8)}}{dt^2}-\frac 1t\frac{d\phi
^{(8)}}{dt}-\frac{g^2v^2\rho^2}{4t^3}\phi ^{(8)}(t)
\nonumber\\&&= 
\frac 1t\frac{d}{dt}\frac{\rho ^2}{2t^3}(\frac{g}{\sqrt
2}A^{(2)}f^{(6)}(t)+\frac{3t^2}{\rho ^2}\frac{d\alpha ^{(2)}(t)}{dt}\phi
^{(6)}(t))
\nonumber\\&&   
+\frac{gv\rho ^2}{\sqrt 2}\frac{1}{4t^5}(-g\sqrt
2A^{(2)}(t)\phi ^{(6)}(t)+4t\frac{d\alpha ^{(2)}(t)}{dt}f^{(6)}(t))
\label{guyellen} 
\end{eqnarray}
and
\begin{equation}
-2t\frac{d\tilde{\psi}^{(8)}(t)}{dt}+4\tilde{\psi}^{(8)}(t)
=3t\frac{d\alpha
^{(2)}(t)}{dt}\phi ^{(6)}(t)
\label{visacard}
\end{equation}
with $\tilde{\psi }^{(8)}(t)=\psi ^{(8)}(t)+\phi ^{(8)}(t)$. (\ref{louellen}) and
(\ref{guyellen}) are special cases of (\ref{crocodyl}), with $n=3$ and
$n=2$, respectively, and their solutions are found from (\ref{minotaur}),
while (\ref{visacard}) is solved by quadrature.  All the terms on the
right-hand sides of (\ref{louellen}), (\ref{guyellen}) and
(\ref{visacard}) have exponential decrease for
$t\rightarrow 0$. This property is, as is demonstrated below,  
shared by the solutions.

The solution of (\ref{visacard}) is
\begin{equation}
\tilde{\psi }^{(8)}(t)=-\frac 32t^2\int _0^t\frac{dt'}{(t')^2}
\frac{d\alpha ^{(2)}(t')}{dt'}\phi ^{(6)}(t')
\label{pterosaur}
\end{equation}
where by (\ref{ichtysaur}) we have $\phi ^{(6)}(t)=O(t^2)$ in lowest order of the expansion
in $v$, and inserting lowest order terms in (\ref{pterosaur}) we find
\begin{equation}
\tilde{\psi }^{(8)}(t)=-\frac{3}{16}g\sqrt 2v\rho ^2t^3+\cdots
\end{equation}
in agreement with (\ref{ursula}). The integrand of (\ref{pterosaur}) shows exponential
decrease for $t\rightarrow 0$ after summation over all orders of $v$, and
the integral therefore also does so.

To the lowest order of $v$ the right-hand side of (\ref{louellen}) is
$O(t)$ and of (\ref{guyellen}) it is $O(t^{-1})$ while the solutions are
$O(t^4)$ and $O(t^3)$. 
In the integral of (\ref{minotaur}):
\begin{equation}
\int
_{t_0}^tdt'(f_{1}(t)f_{2}(t')-f_{1}(t')f_{2}(t))W^{-1}(t')J(t')
\label{mesotaur}
\end{equation}
the estimates $f_1(t)=O(t^0)$, $f_2(t)=O(t^{n})$ following from
(\ref{Bessel}), (\ref{whitaker}) and (\ref{watson}) for $n>0$ are used. For
(\ref{louellen}) one finds
\begin{eqnarray}&&
\int _0^tdt'f_2(t')W^{-1}(t')J(t')
\nonumber\\&&=\int_0^tdt'O((t')^3)O((t')^{-2})
O(t')=O(t^3)  
\end{eqnarray}
and
\begin{eqnarray}&&
\int
_{t_0}^tdt'f_1(t')W^{-1}(t')J(t')
\nonumber\\&&=\int_{t_0}^tdt'O((t')^0)O((t')^{-2}),
O(t')=O(t^{0})  
\end{eqnarray}
respectively, so (\ref{mesotaur}) is $O(t^3)$. For (\ref{guyellen}) the results are
\begin{eqnarray}&&
\int _{0}^tdt'f_2(t')W^{-1}(t')J(t')
\nonumber\\&&=\int_0^tdt'O((t')^2)O((t')^{-1})
O((t')^{-1})=O(t)  
\end{eqnarray}
and
\begin{eqnarray}&&
\int
_{t_0}^tdt'f_1(t')W^{-1}(t')J(t')
\nonumber\\&&=\int_{t_0}^tdt'O((t')^0)O((t')^{-1})
O((t')^{-1})=O(t^{-1})
\end{eqnarray}
so here (\ref{mesotaur}) is $O(t)$. 
Thus in neither case is there is any possibility of
having a term proportional to $f_1(t)$ in the solution, and
asymptotic falloff
is ensured.

As in the bosonic case this iteration procedure can be continued and the
conclusions carry over to higher orders.

\section{The
superconformal zero mode}
\setcounter{equation}{0}

The equations (\ref{lambada})-(\ref{tildequ}) also have
as a solution the superconformal zero mode.
In this case   the fields have the forms:
\begin{equation}
\lambda_B ^a=f(t)\bar{\sigma }\cdot x\sigma ^au_\sigma,
\end{equation}
\begin{equation}
q_A=i\phi(t)\tau ^ai\tau _2u_\tau\otimes \sigma ^ au_\sigma
+i\psi(t)i\tau 2u_\tau
\otimes u_\sigma
\end{equation}
and
\begin{equation}
\tilde{q}_A=i\phi(t)\tau ^au_\tau\otimes \sigma
_au_\sigma +i\psi(t)u_\tau\otimes u_\sigma.
\end{equation}
 Then the
following set of  coupled equations is obtained by the identities of App.
A:
\begin{equation}
4f(t)-2t\frac{df(t)}{dt}
-4t\frac{d\log \alpha (t)}{dt}f(t)=-g\sqrt 2A(t)\phi(t),
\label{corqueran} 
\end{equation}
\begin{equation}
2\frac{d\psi (t)}{dt}+3\frac{d\log \alpha (t)}{dt}\phi (t)=0
\label{bigben}
\end{equation}
and
\begin{equation}
2\frac{d\phi (t)}{dt}+\frac{d\log \alpha (t)}{dt}
(\psi (t)-2\phi (t))=g\sqrt
2\frac{\rho^2}{2t^2}A(t)f(t).
\label{westminster}
\end{equation}

\subsection{Iteration up to first order}

At zeroth order the solution of
(\ref{corqueran}) is:
\begin{equation}
f_0(t)=\frac{t^2}{(1+t)^2}.\label{SCO}
\end{equation}
At first order
(\ref{bigben}) and (\ref{westminster}) are:
\begin{equation}
2\frac{d\psi _1(t)}{dt}+\frac{3\phi
_1(t)}{1+t}=0
\label{smallben}
\end{equation}
and
\begin{equation}
2\frac{d\phi _1(t)}{dt}+\frac{\psi _1(t)-2\phi _1(t)}{1+t}=\frac{g\sqrt
2v\rho ^2}{2}\frac{1}{(1+t)^\frac 5 2}
\label{eastminster}
\end{equation}
with the solutions:
\begin{equation}
\phi _1(t)=\frac
14\frac{C_1}{(1+t)^\frac 12}+\frac 34C_2(1+t)^{\frac 32}-\frac{g\sqrt 2v\rho
^2}{8}\frac{1}{(1+t)^\frac 32}
\label{quiggin}
\end{equation}
and
\begin{equation}
\psi _1(t)=\frac 34\frac{C_1}{(1+t)^\frac 12}-\frac 34C_2(1+t)^{\frac
32}-\frac{g\sqrt 2v\rho ^2}{8}\frac{1}{(1+t)^\frac 32}
\label{princeling}
\end{equation}
where $C_1$ and $C_2$ are integration constants.

A solution that vanishes at $t=0$ is obtained by the choice:
\begin{equation}
C_1=\frac{g\sqrt 2v\rho^2}{4};\hspace{1 mm}C_2=\frac{g\sqrt 2v\rho^2}{12}
\end{equation}
i.e.
\begin{equation}
\phi _1(t)=\frac{g\sqrt 2v\rho
^2}{8}(\frac
12\frac{1}{(1+t)^\frac 12}+\frac 12(1+t)^{\frac 32}
-\frac{1}{(1+t)^\frac 32})
\label{Husar}
\end{equation}
and
\begin{equation}
\psi _1(t)=\frac{g\sqrt 2v\rho
^2}{8}(\frac 32\frac{1}{(1+t)^\frac 12}-\frac 12(1+t)^{\frac
32}-\frac{1}{(1+t)^\frac 32}).
\label{Victoria}
\end{equation}
However, this solution is singular at the origin; a solution that is
regular at the origin (for
$t\rightarrow \infty$) can only be obtained for $C_2=0$.

It was suggested \cite{ADS} that a regular superconformal zero mode
can be obtained by introducing additional Yukawa coupling terms in the
Lagrangian through a constraint. Several possibilities for doing so
suggest themselves.  

The terms that grow too fast for $t\rightarrow \infty$ in
(\ref{Husar}) and (\ref{Victoria}) are eliminated by adding to the
solution: 
\begin{equation}
\phi _{1, \rm {add}}(t)=-\frac{g\sqrt 2v\rho
^2}{16}\frac{2t+t^2}{\sqrt{1+t}}
\label{tonorm}
\end{equation}
and
\begin{equation}
\psi _{1, \rm {add}}(t)=\frac{g\sqrt 2v\rho ^2}{16}
\frac{2t+t^2}{\sqrt{1+t}}.
\label{enorm}
\end{equation}
This means replacing (\ref{qu}) by:
\begin{eqnarray}&&
\bar{\sigma }\cdot D q_A+ig\sqrt{2}\frac{\tau
^a}{2}\lambda _B^ai\tau ^2A^*
\nonumber\\&&
=i\frac{t^2}{\rho ^2}\frac{g\sqrt
2v\rho ^2}{4}(1+t)^{-\frac 32} (\tau ^au_\tau\otimes \bar{\sigma }\cdot
x\sigma ^au_\sigma 
\nonumber\\&&
-u_\tau\otimes \bar{\sigma }\cdot xu_\sigma).
\end{eqnarray}
In order to produce  the right-hand side by a Yukawa coupling one
evidently needs an isoscalar field in addition to the isovector gluino field
$\lambda_B ^a$. This feature persists  when
in (\ref{tonorm}) and (\ref{enorm}) the numerator
$2t+t^2$ is replaced by a different function of $t$.

Keeping instead $C_2=0$ but choosing:
\begin{equation}
C_1=\frac{g\sqrt 2v\rho ^2}{2}
\label{zoroaster}
\end{equation}
one obtains that $\phi _1(t)$ vanishes for $t=0$, and 
(\ref{quiggin}) and (\ref{princeling})  are:
\begin{equation}
\phi _1(t)=\frac{g\sqrt 2v\rho
^2}{8}(\frac{1}{(1+t)^\frac 12}-\frac{1}{(1+t)^\frac 32})
\label{pershing}
\end{equation}
and
\begin{equation}
\psi _1(t)=\frac{g\sqrt 2v\rho
^2}{8}(\frac{3}{(1+t)^\frac 12}-\frac{1}{(1+t)^\frac 32}).
\label{madison}
\end{equation}

We can instead of (\ref{madison})  postulate a solution
obtained by addition  of
\begin{equation}
\psi _{1, \rm add}(t)=-\frac{g\sqrt 2v\rho ^2}{4}
\end{equation}
thus obtaining also  $\psi_1(0)=0$. 
This is achieved by modifying (\ref{qu}) into
\begin{eqnarray}&&
\bar{\sigma}\cdot Dq_A+ig\sqrt 2\frac{\tau ^a}{2}\lambda
^a_Bi\tau ^2A^*
\nonumber\\&&
=i\frac{g\sqrt 2v}{4}\frac{t^2}{1+t}\tau
^a i\tau ^2u_\tau\otimes
\bar{\sigma }\cdot x\sigma ^au_\sigma
\nonumber\\&&\simeq i\frac{g\sqrt 2v}{4}(1+t)\tau ^a i\tau ^2u_\tau
\otimes
\lambda ^a_B,
\label{hoppetossa}
\end{eqnarray} 
where no isoscalar term occurs, and
(\ref{tildequ}) should be modified correspondingly.
Thus the following new Yukawa coupling terms should be present in
(\ref{lagslut}):
\begin{equation}
-i\frac{g\sqrt 2v}{4}(1+t)(q_B^\dagger\tau ^a i\tau ^2u_\tau 
+\tilde{q}_B^\dagger\tau ^a u_\tau)\otimes
\lambda ^a_B.
\label{Barbiroussa}
\end{equation}

\subsection{Ultraviolet and infrared asymptotic behaviour}

For $t\rightarrow \infty $ one has from (\ref{corqueran}), (\ref{bigben}) and
(\ref{westminster}) and from the low order computations  the following
consistent estimates:
\begin{equation}
f_0\rightarrow 1; \hspace{1 mm}f_n\propto t^{-1},\hspace{1 mm}n \neq 0; 
\phi \propto
t^{-\frac 32}, \phi -\psi \propto t^{-\frac 52}.
\end{equation}

For $t\rightarrow 0$  leading, nextleading etc. terms are found by a
systematic expansion in $\rho$ and
$\sqrt t$, starting in second order, where  the solutions
(\ref{pershing})-(\ref{madison}) imply
\begin{equation}
\psi ^{(2)}=\frac{g\sqrt 2v\rho ^2}{4}+\cdots
\label{minnetonka}
\end{equation}
while in fourth order
\begin{eqnarray}&&
f^{(4)}(t)=t^2+\cdots,\hspace{1 mm}\psi ^{(4)}(t)=O(t^2), \nonumber\\&&\phi
^{(4)}(t)=\frac{g\sqrt 2v\rho ^2}{8}t+\cdots.
\end{eqnarray}
The equations (\ref{corqueran}), (\ref{bigben}) and (\ref{westminster}) require in
second order that $\psi ^{(2)}$ is indeed a constant while the nextleading
terms are analyzed in the same way and with the same conclusions as
for the supersymmetric zero mode.

\section{The quark zero mode}

The quark zero mode is according to (\ref{lagslut}) determined by:
\begin{eqnarray} &&
\sigma \cdot D q_B-ig\sqrt{2}\frac{\tau
^a}{2}\lambda ^a_AA=0,  \label{ququ}\\&&
\sigma \cdot D \tilde{q}_B-ig\sqrt{2}\frac{\tau
^a}{2}\lambda ^a_A\tilde{A}=0 \label{tildeququ}
\end{eqnarray}
and
\begin{equation}
(\bar{\sigma }\cdot
D )^{ab}\lambda ^b_A+ig\sqrt{2}A^\dagger \frac{\tau ^a}{2}q_B
+ig\sqrt{2}\tilde{A}^\dagger \frac{\tau
^a}{2}\tilde{q}_B=0.  \label{lambadaqu}
\end{equation}
In 
(\ref{ququ})-(\ref{lambadaqu}) the following Ans\"{a}tze are used:
\begin{equation}
q_B(x)=\phi(t)\tau ^au_\tau\otimes\bar{\sigma }\cdot
x\sigma ^au_\sigma+\psi(t)u_\tau\otimes\bar{\sigma }\cdot
xu_\sigma ,
\end{equation}
\begin{eqnarray}&&
\tilde{q}_B(x)=\phi(t)\tau ^a(-i\tau ^2)u_\tau\otimes\bar{\sigma }\cdot
x\sigma ^au_\sigma
\nonumber\\&&+\psi(t)(-i\tau ^2)u_\tau\otimes\bar{\sigma }\cdot
xu_\sigma
\end{eqnarray}
and
\begin{equation}
\lambda _A^a(x)=if(t)\sigma ^au_\sigma \label{dareios}
\end{equation}
and one obtains the three equations:
\begin{eqnarray}&&
4\phi(t)-2t\frac{d\phi (t)}{dt}-2t\frac{d\log \alpha
(t)}{dt}\phi (t)+t\frac{d\log \alpha(t)}{dt}\psi
(t)
\nonumber\\&&=-\frac{g}{\sqrt 2}A(t)f(t),
\label{acanthus}
\end{eqnarray}
\begin{equation}
4\psi (t)-2t\frac{d\psi (t)}{dt}+3t\frac{d\log \alpha
(t)}{dt}\phi (t)=0
\label{cactus}
\end{equation}
and\begin{equation}
-\frac{2t^2}{\rho
^2}(\frac{df(t)}{dt}-2\frac{d\log
\alpha (t)}{dt}f(t))=-g\sqrt{2}A(t)\phi(t).
\label{smerdis}
\end{equation}

To zeroth order the solutions are 
\begin{equation}
\phi _0(t)=-\psi _0(t)=
\frac{t^2}{(1+t)^{\frac 32}}
\end{equation}
whence the following first order solution of (\ref{smerdis}) is found
\begin{equation}
f_1(t)=-\frac{g\sqrt{2}v\rho ^2}{6}\frac{1}{1+t}.
\label{balaleika}
\end{equation}

Adding to $f(t)$:
\begin{equation}
f_{1, {\rm add}}(t)=\frac{g\sqrt{2}v\rho ^2}{6}
\end{equation}
one obtains $f(0)=0$. This is achieved by the new Yukawa coupling
terms in (\ref{lagslut}):
\begin{eqnarray}&&
-\frac 23igv\sqrt{2}(1+t)^{\frac 12}(\lambda
^a_B)^\dagger (x)(u^\dagger_\tau \frac{\tau
^a}{2}q_B(x)
\nonumber\\&&
+u^\dagger_\tau i\tau ^2
\frac{\tau ^a}{2}\tilde{q}_B(x)).
\end{eqnarray}
They are similar to (\ref{Barbiroussa}) but with a different
coefficient.
Considerations on matters of asymptotic behaviour are very similar both in
methods and results to the analysis presented above for the superconformal
zero mode.

\section{Conclusion}

In this paper a systematic way of dealing with mass corrections of
fermionic zero modes in the presence of constrained instantons was
presented, following the general pattern applied to the constrained
instanton itself in \cite{NN}.  Low order iterations  as
well as general estimates of the asymptotic behaviour both at long and
short distances were presented. While the supersymmetric zero mode was
found to be well behaved, the superconformal and quark zero modes go at
large distances as  constants at first order in $v$. 

However it was found, following the
suggestion of \cite{ADS}, that these unpermissible first order terms can be
removed by new Yukawa couplings, which consequently are of first order in
$v$, in contrast to the gauge constraint terms, which are of second and
fourth order in $v$.

No operator constraint was determined. As pointed out in \cite{NN}, the
existence of a gauge invariant operator constraint is unlikely, and even
if it exists, it can probably not be implemented in a supersymmetric
fashion because of the mismatch of the powers of $v$ in the bosonic and
the fermionic sector.

\appendix 
\setcounter{equation}{0}

\section{Algebraic identities} \label{Ydyntytyt}

A number of useful algebraic identities involving the
matrices $\sigma _\mu =(-i\vec{\sigma},1)$ and
$\bar{\sigma}_\mu=(i\vec{\sigma},1)$ is recorded:
\begin{equation}
\sigma _\mu \bar{\sigma }_\nu+\sigma _\nu \bar{\sigma }_\mu
=\bar{\sigma }_\mu \sigma _\nu+\bar{\sigma} _\nu \sigma
_\mu=2\delta _{\mu \nu },
\end{equation}
\begin{equation}
\bar{\eta }^a_{\mu \nu }\sigma _\mu x_\nu =i\sigma ^a\sigma
\cdot x;\hspace{1 mm}\bar{\eta }^a_{\mu \nu }\bar{\sigma }_\mu x_\nu=-i\bar{\sigma }\cdot
x\sigma ^a ,\label{kedelig}
\end{equation}
\begin{equation}
\bar{\sigma }_\mu \sigma ^a\sigma _\mu =0,
\end{equation}
\begin{eqnarray}
&&\tau _\lambda \bar{\tau
}_\mu u_\tau \otimes \bar{\sigma }_\lambda \sigma _\mu
u_\sigma =4u_\tau
\otimes u_\sigma,
\end{eqnarray}
\begin{eqnarray}
&&\tau \cdot x\bar{\tau }_\lambda u_\tau \otimes \bar{\sigma
}\cdot x
\sigma _\lambda u_\sigma 
\nonumber\\&&=x^2u_\tau
\otimes u_\sigma+\tau ^au_\tau \otimes \bar{\sigma }\cdot
x\sigma ^a\sigma \cdot x u_\sigma
\label{iddentitet}
\end{eqnarray}
and
\begin{eqnarray}
&&i\bar{\eta
}^a_{\mu \nu }x_\nu \tau ^a \tau \cdot x\bar{\tau
}_\lambda u_\tau \otimes \bar{\sigma }_\mu \sigma _\lambda
u_\sigma 
\nonumber\\&&
=4x^2u_\tau
\otimes u_\sigma-\tau \cdot x\bar{\tau }_\lambda u_\tau \otimes \bar{\sigma
}\cdot x
\sigma _\lambda u_\sigma .  
\end{eqnarray}

\section{Bessel functions}
\setcounter{equation}{0}

The modified Bessel equation \cite{Watson} of order $n $ 
\begin{equation}
z^2\frac{d^2G_n(z)}{dz^2}+z\frac{dG_n(z)}{dz}-(z^2+n ^2)G_n(z)=0
\label{Bessel}
\end{equation}
has two linearly independent solutions $I_n (z)$ and $K_n (z)$.
For $z\rightarrow \infty$:
\begin{equation}
I_n (z)\simeq \sqrt{\frac{1}{2\pi x}}e^{z};\hspace{1 mm}
K_n (z)\simeq \sqrt{\frac{\pi }{x}}e^{-z}.
\label{ask}
\end{equation}

The power series expansions of the modified Bessel functions are
\begin{equation}
I_n(z)=(\frac z2)^n\sum
_{r=0}^\infty\frac{(\frac{z^2}{4})^r}{r!(n+r)!}
\label{whitaker}
\end{equation}
and
\begin{eqnarray}&&
K_n(z)=(-1)^{n+1}I_n(z)(\log \frac z2+\gamma )
\nonumber\\&&+\frac 12\sum
_{r=0}^{n-1}\frac{(-1)^r(n-r-1)!}{r!}(\frac z2)^{-n+2r}
\nonumber\\&&
+\frac 12(-\frac z2)^n\sum
_{r=0}^\infty\frac{1}{r!(n+r)!}(\phi(r)+\phi (n+r))(\frac
{z^2}{4})^{r}
\label{watson}
\end{eqnarray}
with
\begin{equation}
\phi(0)=0,\hspace{1 mm}\phi(r)=\sum _{s=1}^r\frac 1s, r\neq 0.
\end{equation}

From the definition $t=\frac{\rho ^2}{x^2}$  follows:
\begin{eqnarray}
&&(\frac{d^2}{dt^2}+\frac{1-n }{t}\frac{d}{dt}-\frac{\rho
^2m^2}{4t^3})(\frac{m}{x})^{n} G_{n}(mx)
\nonumber\\&&
=\frac{1}{x^2}\frac{\rho ^2}{4t^3}(\frac{m}{x})^{n}
[m^2x^2G_n''(mx )+mxG_n'(mx) 
\nonumber\\&&-(m^2x^2+n^2)G(mx)].
\end{eqnarray}
Here the expression in the square bracket is recognized as the defining
equation of the modified Bessel functions (\ref{Bessel}) of order $n$.
Consequently the solutions of the equation
\begin{equation}
(\frac{d^2}{dt^2}+\frac{1-n }{t}\frac{d}{dt}-\frac{\rho
^2m^2}{4t^3})f(t)=0
\label{aliceford}
\end{equation}
 are
\begin{equation}
f_1(t)=(\frac{m}{x})^{n} I_{n}(mx)\hspace{1 mm}{\rm or}\hspace{1 mm}
f_2(t)=(\frac{m}{x})^{n}
K_{n}(mx).\label{Bessel}
\end{equation}

In terms of $t$ and using the estimates (\ref{ask}) 
valid near $t=0$ the Wronskian $W(t)$ of the two solutions is:
\begin{equation}
W(t)=f_{1}(t)\frac{df_{2}(t)}{dt}-f_{2}(t)\frac{df_{1}(t)}{dt}\simeq 
\frac{1}{\sqrt 2}(\frac{m}{\rho })^{2n}t^{n-1}.
\label{ratata}
\end{equation}
It fulfils:
\begin{equation}
\frac{dW(t)}{dt}=\frac{n-1}{t}W(t).
\end{equation}
Thus (\ref{ratata}), which was derived only for $t\simeq 0$, actually is valid as an
equality for all values of $t$.

The general inhomogeneous version of (\ref{aliceford}) is (\ref{crocodyl})
with the general solution (\ref{minotaur}).
The integral here is near $t=0$  by (\ref{ask}) and (\ref{ratata})
proportional to
\begin{equation}
-2\int ^tdt'(\sqrt {tt'})^{n}\frac{^4\sqrt {tt'}}{m\rho
}(t')^{1-n}\sinh
\rho(\frac{1}{\sqrt t}-\frac{1}{\sqrt t'})J(t').
\label{maxotaur}
\end{equation}
If $J(t)$ falls off exponentially for $t\rightarrow  0$ it is
immediately clear from (\ref{minotaur}) or (\ref{maxotaur}) that the
same holds for
$f(t)$.


\begin{thebibliography}{00}
\bibitem{Affleck} I. Affleck, Nucl. Phys. {\bf B 191}, 429 (1981).
\bibitem{NN} M. Nielsen, N.K. Nielsen,  Phys Rev. {\bf  D61}, 105020
(2000).
\bibitem{ADS}
I. Affleck, M. Dine, N. Seiberg,  Nucl. Phys. {\bf B 241}, 493
(1984).
\bibitem{NKN} N.K.Nielsen, "Instanton constraints in supersymmetric gauge
theories II. $N=2$ Yang-Mills theory", hep-th/053120.
\bibitem{WB} J. Wess, J.Bagger, Supersymmetry and supergravity, 2nd ed.
Princeton University Press, 1992
\bibitem{Vainshtein}  A.I. Vainshtein, V.I. Zakharov, JETP Lett. {\bf 35}, 
323 (1982).
\bibitem{Espinosa} O. Espinosa, Nucl. Phys. {\bf B343}, 310 (1990).
\bibitem{Watson} G.N.Watson, Theory of Bessel functions, Cambridge
University Press, 1944.
\end{thebibliography}
\end{document}